\documentclass[
twocolumn, 
superscriptaddress,
 showpacs,preprintnumbers,
 amsmath,amssymb,
 aps,
 prb,
]{revtex4-1}
\usepackage{graphicx}
\usepackage{dcolumn}
\usepackage{bm}

\begin{document}


\title{Non-stationary effects in the system of coupled quantum dots influenced by the Coulomb correlations}

\author{P.\,I.\,Arseyev}
 \altaffiliation{ars@lpi.ru}
\author{N.\,S.\,Maslova}%
 \email{spm@spmlab.phys.msu.ru}
\author{V.\,N.\,Mantsevich}
 \altaffiliation{vmantsev@spmlab.phys.msu.ru}
\affiliation{%
 P.N. Lebedev Physical institute of RAS, 119991, Moscow, Russia\\~\\
 Moscow State University, Department of  Physics,
119991 Moscow, Russia
}%

\date{\today }

\begin{abstract}
We found an analytical solution for the time dependent filling
numbers of the localized electrons in a system of two coupled
single-level quantum dots (QDs) connected with continuous spectrum
states in the presence of Coulomb interaction. This solution takes
into account correlation functions of all orders for the electrons
in the QDs by decoupling high order correlations between localized
and band electrons.

We demonstrated that several time scales with the strongly different
relaxation rates appear in the system for a wide range of the
Coulomb interaction value. We found that specific non-monotonic
behavior of charge relaxation in QDs takes place due to Coulomb
correlations.

We also found that besides the usual charge oscillations with the
period determined by the detuning between the QDs energy levels a
new effect of period doubling appears in the presence of Coulomb
interaction at particular range of the system parameters.
\end{abstract}

\maketitle

\section{Introduction}

The control and manipulation of localized charge in the small size
systems is one of the most important points in nanoelectronics.
\cite{Collier, Gittins} Single semiconductor QDs which are referred
as "artificial" atoms \cite{Kastner, Ashoori} and coupled QDs -
"artificial" molecules \cite{Oosterkamp, Blick_0} are perspective
structures that may serve for creation of extremely small devices.
Several coupled QDs can be used for electronic devices creation
dealing with quantum kinetics of individual localized states.
\cite{Stafford_0, Hazelzet, Cota} Due to this fact the behavior of
coupled QDs in different configurations is recently under careful
experimental \cite{Waugh, Blick} and theoretical investigation.
\cite{Stafford, Matveev}

During the last decade vertically aligned QDs have been fabricated
and widely studied with the great success (for example indium
arsenide QDs in gallium arsenide).\cite{Vamivakas, Stinaff,
Elzerman} Such experimental realization allows to organize strongly
interacting QDs system with only one of them coupled to the
continuous spectrum states. Consequently vertically aligned QDs give
an opportunity to analyze non-stationary effects in various charge
and spin configurations formation in the small size structures.
\cite{Kikoin}

Lateral QDs seems to be better candidates for controllable
electronic coupling between two or several QDs by applying
individual lateral gates. That's why they are intensively studied
during the last several years both experimentally and theoretically.
\cite{Peng, Munoz-Matutano}

Investigation of relaxation processes, non-equilibrium charge
distribution  and non-stationary effects in the electron transport
through the system of QDs are vital problems which should be solved
to integrate QDs in small quantum circuits. \cite{Angus,
Grove-Rasmussen, Moriyama, Landauer, Loss, Nigg, Filippone} Electron
transport in such systems is strongly governed by the Coulomb
interaction between localized electrons and of course by the ratio
between the tunneling transfer amplitudes and the QDs coupling.
Correct interpretation of quantum effects in nanoscale systems gives
an opportunity to create high speed electronic and logic devices.
\cite{Tan, Hollenberg} In some of the recent realizations the
Coulomb interaction is weak, \cite{Feve} but for small size QDs the
on-site Coulomb repulsion is in general strong, \cite{López}
consequently it is important to take it into account. In some cases
Coulomb correlations can determine time-dependent phenomena.
\cite{Reckermann} So the problem of time evolution of the charge in
the coupled QDs connected with the continuous spectrum states in the
presence of Coulomb correlations between the localized electrons is
really vital.

Time evolution of charge states in the semiconductor double quantum
well in the presence of Coulomb interaction was experimentally
studied in.\cite{Hayashi} The authors manipulated the localized
charge by the initial pulses and observed pulse-induced tunneling
electrons oscillations. Time dependence of the accumulated charge
and the tunneling current through the single QD in the presence of
Coulomb interaction was theoretically analyzed in. \cite{Contreras}
The authors described relaxation processes and revealed three time
rates for localized charge relaxation in the QD coupled with the
thermostat. Several different time rates were also found in the
system of two and three interacting QDs coupled with the reservoir.
\cite{Pump, Mantsevich_1, Mantsevich}

In this paper we consider charge relaxation in the double QDs due to
the coupling with the continuous spectrum states. Tunneling from the
first QD to the continuum is possible only through  the second dot.
We obtained the closed system of equations for time evolution of the
localized electrons filling numbers which exactly takes into account
all order correlation functions for localized electrons. It allows
to find an exact analytical solution for the time dependent filling
numbers of the electrons by decoupling the high order correlation
functions between conduction electrons in the reservoir (band
electrons) and electrons localized in the QDs. In such an
approximation the electrons distribution in the reservoir is not
influenced by changing of the electronic states in the coupled QDs.
For QDs weakly coupled to the reservoir the proposed decoupling
scheme is a good approximation. We found some peculiarities in
filling numbers for the electrons dynamics arising due to the
Coulomb correlation effects.

\section{Model}

We consider a system of coupled QDs with the single particle levels
$\varepsilon_1$ and $\varepsilon_2$ connected to an electronic
reservoir (Fig. \ref{figure_1}). At the initial time two electrons
with opposite spins are localized in the first QD on the energy
level $\varepsilon_1$ ($n_{1\sigma}(0)=n_{0}=1$). The second QD with
the energy level $\varepsilon_2$ is connected with the continuous
spectrum states ($\varepsilon_p$). Relaxation of the localized
charge is governed by the Hamiltonian:

\begin{figure} [t]
\includegraphics[width=60mm]{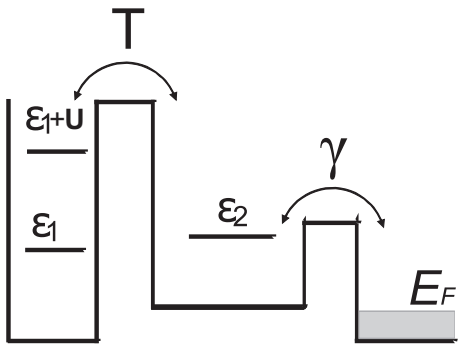}
\caption{Scheme of the proposed model. The system of interacting QDs
is coupled with the continuous spectrum states by means of the
tunneling rate $\gamma=\pi\nu_0t^{2}$.} \label{figure_1}
\end{figure}

\begin{eqnarray}
\Hat{H}=\Hat{H}_{D}+\Hat{H}_{tun}+\Hat{H}_{res}.
\end{eqnarray}

The Hamiltonian $\Hat{H_{D}}$ of interacting QDs

\begin{eqnarray}
\Hat{H}_{D}&=&\sum_{i=1,2\sigma}\varepsilon_{i}c^{+}_{i\sigma}c_{i\sigma}+Un_{1\sigma}n_{1-\sigma}\nonumber\\&+&\sum_{\sigma}T(c_{1\sigma}^{+}c_{2\sigma}+c_{1\sigma}c_{2\sigma}^{+}),
\end{eqnarray}

contains the spin-degenerate levels $\varepsilon_i$ (indexes $i=1$
and $i=2$ correspond to the first and to the second QD) and the
on-site Coulomb repulsion for the double occupation of the first
dot. For simplicity we consider Coulomb interaction only in the
first QD though it is possible to obtain closed system of equations
for filling numbers correlators in a general case taking into
account Coulomb interaction between all the electrons localized in
the dots. Our model is suitable for the case when the first QD is
narrow and the second one is rather wide.\cite{Mantsevich_1,
Kikoin_1} Besides, if electrons are initially located in the first
QD and the second dot is empty, then filling numbers for the
electrons in the second QD remain rather small during the time
evolution of the charge and Coulomb effects in the second QD are not
so important as in the first one.

The creation/annihilation of an electron with spin $\sigma=\pm1$
within the dot is denoted by $c^{+}_{i\sigma}/c_{i\sigma}$ and
$n_{\sigma}$ is the corresponding filling number operator. The
coupling between the dots is described by the tunneling transfer
amplitude $T$ which is considered to be independent of momentum and
spin.

The continuous spectrum states are modeled by the Hamiltonian:

\begin{eqnarray}
\Hat{H}_{res}=\sum_{p\sigma}\varepsilon_{p}c^{+}_{p\sigma}c_{p\sigma},
\end{eqnarray}

where $c^{+}_{p\sigma}/c_{p\sigma}$ creates/annihilates an electron
with spin $\sigma$ and momentum $p$ in the lead. The coupling
between the second dot and the continuous spectrum states is
described by the Hamiltonian:

\begin{eqnarray}
\Hat{H}_{tun}=\sum_{p\sigma}t(c_{p\sigma}^{+}c_{2\sigma}+c_{p\sigma}c_{2\sigma}^{+}),
\end{eqnarray}

where $t$ is the tunneling amplitude, which we assume to be
independent on momentum and spin. By considering a constant density
of states in the reservoir $\nu_0$, the tunnel rate $\gamma$ is
defined as $\gamma=\pi\nu_0t^{2}$.

As we are interested in the specific features of the non-stationary
time evolution of the initially localized charge in the coupled QDs,
we'll consider the situation when condition
$(\varepsilon_i-\varepsilon_F)/\gamma>>1$ is fulfilled. It means
that initial energy levels are situated well above the Fermi level
and stationary occupation numbers in the second QD in the absence of
coupling between the QDs is of the order
$\gamma/(\varepsilon_2-\varepsilon_F)<<1$ and can be omitted.
Consequently the Kondo effect is also negligible  in the proposed
model.

Our investigations deal with the low temperature regime when Fermi
level is well defined and the temperature is much lower than all
typical relaxation rates in the system. Consequently the
distribution function of electrons in the leads (band electrons) is
a Fermi step.

We set $\hbar=1$ and therefore the kinetic equations for bilinear
combinations of Heisenberg operators $c_{i\sigma}^{+}/c_{i\sigma}$

\begin{eqnarray}
c_{1\sigma}^{+}c_{1\sigma}=\hat n_{1}^{\sigma}(t);\quad
c_{2\sigma}^{+}c_{2\sigma}=\hat n_{2}^{\sigma}(t);\nonumber\\
c_{1\sigma}^{+}c_{2\sigma}=\hat n_{12}^{\sigma}(t);\quad
c_{2\sigma}^{+}c_{1\sigma}=\hat n_{21}^{\sigma}(t),
\end{eqnarray}

which describe time evolution of the filling numbers for the
electrons can be written as:

\begin{eqnarray}
i\frac{\partial}{\partial
t}\hat n_{1}^{\sigma}&=&-T(\hat n_{21}^{\sigma}-\hat n_{12}^{\sigma}),\nonumber\\
i\frac{\partial}{\partial t}\hat n_{2}^{\sigma}&=&T(\hat
n_{21}^{\sigma}-\hat n_{12}^{\sigma})- 2i\gamma \hat
n_{2}^{\sigma},\nonumber\\
i\frac{\partial}{\partial t}\hat n_{21}^{\sigma}&=&T(\hat
n_{2}^{\sigma}-\hat n_{1}^{\sigma})
-(\xi+U\hat n_{1}^{-\sigma}) \hat n_{21}^{\sigma}-i\gamma \hat n_{21}^{\sigma},\nonumber\\
i\frac{\partial}{\partial t}\hat n_{12}^{\sigma}&=&-T(\hat
n_{2}^{\sigma}-\hat n_{1}^{\sigma})+(\xi+U\hat n_{1}^{-\sigma})
\hat n_{12}^{\sigma}-i\gamma \hat n_{12}^{\sigma},\nonumber\\
\label{system}
\end{eqnarray}

where $\xi=\varepsilon_1-\varepsilon_2$ is the detuning between the
energy levels in the QDs. The system of Eqs. (\ref{system}) contain
expressions for the pair correlators $\hat{n}_{1}^{-\sigma}\hat
n_{21}^{\sigma}$ and $\hat n_{1}^{-\sigma}\hat n_{12}^{\sigma}$,
which also determine relaxation of the localized charge and
consequently have to be evaluated. In this system we neglect high
order correlation functions between localized and continuous
spectrum (band) electrons and fulfill averaging over electron states
in the reservoir.

Let us introduce the following designation for the pair correlators:
$K_{iji^{'}j^{'}}^{\sigma\sigma^{'}}=<c_{i\sigma}^{+}c_{j\sigma}c_{i^{'}\sigma^{'}}^{+}c_{j^{'}\sigma^{'}}>$
and consider only the paramagnetic case $<\hat n_{i}^{\sigma}>=<\hat
n_{i}^{-\sigma}>$. Then the following relations take place

\begin{eqnarray}
K_{2111}^{\sigma-\sigma}&=&<\hat n_{21}^{\sigma}\hat
n_{1}^{-\sigma}>=<\hat n_{21}^{-\sigma}\hat n_{1}^{\sigma}>,\nonumber\\
K_{1211}^{\sigma-\sigma}&=&<\hat n_{12}^{\sigma}\hat
n_{1}^{-\sigma}>=<\hat n_{12}^{-\sigma}\hat n_{1}^{\sigma}>.
\end{eqnarray}

The system of equations for pair correlators can be written in the
compact matrix form (symbol $[\quad]$ means commutation and symbol
$\{\quad\}$- anticommutation):

\begin{eqnarray}
i\frac{\partial}{\partial
t}\widehat{K}=[\widehat{K},\widehat{H}^{'}]+\{\widehat{K},\widehat{\Gamma}\}+\widehat{\Upsilon},
\label{system_compact}
\end{eqnarray}

where $\widehat{K}$ is the pair correlators matrix

\begin{eqnarray}
\widehat{K}=
\begin{pmatrix}
K_{2211}^{\sigma-\sigma} & K_{1211}^{\sigma-\sigma} & K_{2221}^{\sigma-\sigma} & K_{1221}^{\sigma-\sigma}\\
K_{2111}^{\sigma-\sigma} & K_{1111}^{\sigma-\sigma} & K_{2121}^{\sigma-\sigma} & K_{1121}^{\sigma-\sigma}\\
K_{2212}^{\sigma-\sigma} & K_{1212}^{\sigma-\sigma} & K_{2222}^{\sigma-\sigma} & K_{1222}^{\sigma-\sigma}\\
K_{2112}^{\sigma-\sigma} & K_{1112}^{\sigma-\sigma} & K_{2122}^{\sigma-\sigma} & K_{1122}^{\sigma-\sigma}\\
\end{pmatrix}=||K_{ij}||,
\label{correlators}
\end{eqnarray}

matrix $\widehat{H}^{'}$ has the following form

\begin{eqnarray}
\widehat{H}^{'}=
\begin{pmatrix}
0 & T & T & 0\\
T & \xi+U & 0 & T\\
T & 0 & -\xi & T\\
0 & T & T & 0\\
\end{pmatrix},
\end{eqnarray}

and the tunneling coupling matrix $\widehat{\Gamma}$ is denoted as:

\begin{eqnarray}
\widehat{\Gamma}=
\begin{pmatrix}
-i\gamma & 0 & 0 & 0\\
0 & 0 & 0 & 0\\
0 & 0 & -2i\gamma & 0\\
0 & 0 & 0 & -i\gamma\\
\end{pmatrix}.
\end{eqnarray}

One can easily find that Eqs. (\ref{system_compact}) contain
expressions for the high-order correlators
$K_{121122}^{\sigma-\sigma-\sigma}$ and
$K_{211122}^{\sigma-\sigma-\sigma}$. Their contribution can be
easily written in the matrix form $\widehat{\Upsilon}$:

\begin{eqnarray}
\widehat{\Upsilon}=
\begin{pmatrix}
0 & 0 & UK_{211122}^{\sigma-\sigma-\sigma} & 0\\
0 & 0 & 0 & 0\\
UK_{121122}^{\sigma-\sigma-\sigma} & 0 & 0 & UK_{211122}^{\sigma-\sigma-\sigma}\\
0 & 0 & UK_{121122}^{\sigma-\sigma-\sigma} & 0\\
\end{pmatrix}.
\end{eqnarray}

Since the evolution starts from the initial state with two electrons
in the first QD and empty second one, the system of Eqs.
(\ref{system_compact}) for the pair correlators  satisfies the
initial conditions: $K_{1111}^{\sigma-\sigma}(0)=1$;
$K_{2222}^{\sigma-\sigma}(0)=0$;
$K_{iji^{'}j^{'}}^{\sigma-\sigma}(0)=0$ for the other combinations
of indexes $i$, $j$. The high-order correlators
$K_{121122}^{\sigma-\sigma-\sigma}$ and
$K_{211122}^{\sigma-\sigma-\sigma}$ are exactly equal to zero due to
the fact that they are the solution of the linear homogeneous system
of equations with zero initial conditions.

The formal solution of the system for the pair correlators [see Eq.
(\ref{system_compact})] can be written with the help of the
evolution operator. Time evolution of the matrix elements $K_{ij}$
[see Eq. (\ref{correlators})] is given by the expression:

\begin{eqnarray}
K_{ij}(t)=\sum_{mn}(e^{-i\widehat{H}t})_{im}K_{mn}(0)
(e^{i\widehat{H}^+t})_{n j},
\end{eqnarray}

where $\widehat{H}$ is defined as: $
\widehat{H}=\widehat{H}^{'}+\widehat{\Gamma}\nonumber\\
$.

Let us introduce the evolution operator:

\begin{eqnarray}
\Phi_{ij}(t)=(e^{-i\widehat{H}t})_{ij}.
\end{eqnarray}

Consequently, the time evolution of the pair correlators can be
found from the following expressions:

\begin{eqnarray}
K_{2111}^{\sigma-\sigma}&=&(e^{-i\widehat{H}t})_{12}K(0)_{22}(e^{i\widehat{H}^{+}t})_{22}=\Phi_{12}(t)\widetilde{\Phi}_{22}(t),\nonumber\\
K_{1211}^{\sigma-\sigma}&=&(e^{-i\widehat{H}t})_{22}K(0)_{22}(e^{i\widehat{H}^{+}t})_{21}=\Phi_{22}(t)\widetilde{\Phi}_{21}(t).\nonumber\\
\end{eqnarray}
Since $K(0)_{22}$ in the matrix [see Eq. (\ref{correlators})] is
equal to $K_{1111}^{\sigma-\sigma}(0)=1$. The evolution operator
$\widetilde{\Phi}_{22}(t)$ can be obtained from the expression for
the operator $\Phi_{22}(t)$ by the following substitutions:
$t\rightarrow-t$ and $\gamma\rightarrow-\gamma$. Pair correlator
$K_{1211}^{\sigma-\sigma}$ is a complex conjugate of
$K_{2111}^{\sigma-\sigma}$.

Finally the evolution operators $\Phi_{ij}(t)$ are determined by the
equations:

\begin{eqnarray}
\begin{pmatrix}i\frac{\partial}{\partial t}\Phi_{12}(t)\\
i\frac{\partial}{\partial t}\Phi_{22}(t)\\ i\frac{\partial}{\partial t}\Phi_{32}(t)\\ i\frac{\partial}{\partial t}\Phi_{42}(t)\\
\end{pmatrix}=\widehat{H}\cdot\begin{pmatrix}\Phi_{12}(t)\\
\Phi_{22}(t)\\ \Phi_{32}(t)\\ \Phi_{42}(t)\\
\end{pmatrix},
\end{eqnarray}

with the initial conditions:

\begin{eqnarray}
\Phi_{ij}(0)=\delta_{ij}.
\end{eqnarray}

The characteristic equation for the evolution operator
$\Phi_{ij}(t)$ eigenvalues $\lambda_i$ has the form:

\begin{eqnarray}
(H_{11}-\lambda)(H_{22}-\lambda)(H_{33}-\lambda)(H_{44}-\lambda)-T^{2}\nonumber\\
\times[(H_{11}-\lambda)(H_{22}-\lambda)+(H_{11}-\lambda)(H_{33}-\lambda)\nonumber\\
+(H_{33}-\lambda)(H_{44}-\lambda)+(H_{22}-\lambda)(H_{44}-\lambda)]=0,\nonumber\\
\end{eqnarray}

where coefficients $H_{11}$, $H_{22}$, $H_{33}$ and $H_{44}$ are
determined as:

\begin{eqnarray}
H_{11}&=&H_{44}=-i\gamma,\nonumber\\
H_{22}&=&\xi+U,\nonumber\\
H_{33}&=&-\xi-2i\gamma.
\end{eqnarray}

Each eigenvalue $\lambda_i$ determines the corresponding
eigenvector:

\begin{eqnarray}
\psi_i=
\begin{pmatrix}
\alpha_i \\
\beta_i \\
\gamma_i \\
\delta_i \\
\end{pmatrix}.
\end{eqnarray}

In our case it is necessary to obtain expressions for the evolution
operators $\Phi_{12}(t)$ and $\Phi_{22}(t)$ with the initial
conditions $\Phi_{22}(0)=1$ and $\Phi_{ij}(0)=0$.

Solution for the system of equations which determines the functions
$\Phi_{12}(t)$ and $\Phi_{22}(t)$ can be written as:

\begin{eqnarray}
\Phi_{12}(t)=\sum_{i=1}^{4}C_{i}\alpha_{i}e^{-i\lambda_{i}t},\nonumber\\
\Phi_{22}(t)=\sum_{i=1}^{4}C_{i}\beta_{i}e^{-i\lambda_{i}t}.\
\end{eqnarray}

where, constants $C_{i}$ can be obtained from the initial conditions
for the system of equations.

\begin{eqnarray}
\sum_{i}C_{i}\alpha_{i}=0,\nonumber\\
\sum_{i}C_{i}\beta_{i}=1,\nonumber\\
\sum_{i}C_{i}\gamma_{i}=0,\nonumber\\
\sum_{i}C_{i}\delta_{i}=0.\
\end{eqnarray}

\begin{figure*} [t]
\includegraphics[width=170mm]{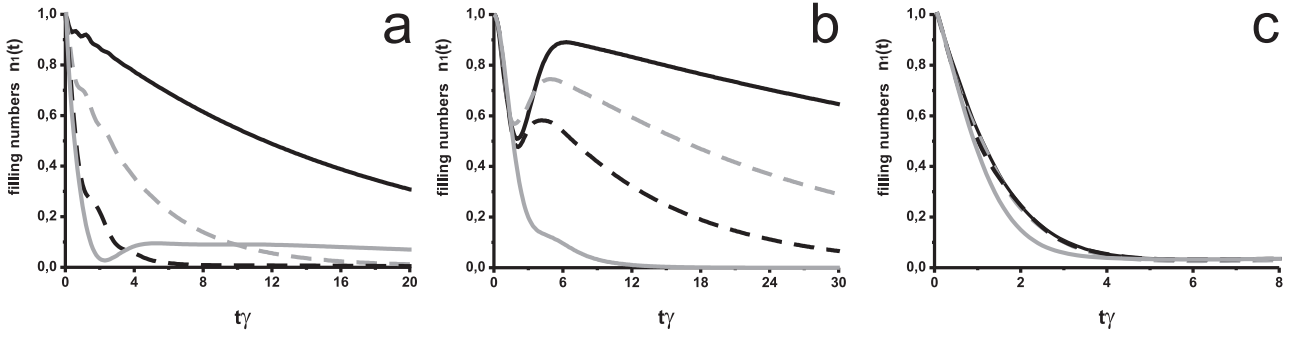}
\caption{Different time evolution regimes of the filling numbers
$n_{1}(t)$ in the first QD in the presence of Coulomb interaction.
a). $(\xi+U)/\gamma=0$ ($U/\gamma=10$, $\xi/\gamma=-10$- black line;
$U/\gamma=5$, $\xi/\gamma=-5$- grey dashed line; $U/\gamma=3$,
$\xi/\gamma=-3$- black dashed line; $U/\gamma=1$, $\xi/\gamma=-1$-
grey line); b). $\frac{\xi+U}{\gamma}\sim1$ ($U/\gamma=10$,
$\xi/\gamma=-7$- black line; $U/\gamma=5$, $\xi/\gamma=-4$- grey
dashed line; $U/\gamma=3$, $\xi/\gamma=-2.5$- black dashed line;
$U/\gamma=1$, $\xi/\gamma=-0.75$- grey line); c). $\xi/\gamma=0$
($U/\gamma=10$- black line; $U/\gamma=5$- grey dashed line;
$U/\gamma=3$- black dashed line; $U/\gamma=1$- grey line).
Parameters $T/\gamma=0.6$, $\gamma=1$ are the same for all the
figures.} \label{figure_2a_2c}
\end{figure*}

\subsection{Equations for the time dependent filling numbers}

The time dependent filling numbers $n_{1}(t)$ can be found from the
inhomogeneous part of Eqs. (\ref{system}), which results in:

\begin{eqnarray}
\{[(i\frac{\partial}{\partial
t}+i\gamma)^{2}+\gamma^{2}][(i\frac{\partial}{\partial
t}+i\gamma)^{2}-\xi^{2}]\nonumber\\-4T^{2}(i\frac{\partial}{\partial
t}+i\gamma)^{2}\}n_{1}(t)=(i\frac{\partial}{\partial
t}+2i\gamma)\nonumber\\ \times T
U(G_{2}^{-1}K_{2111}^{\sigma-\sigma}+G_{1}^{-1}K_{1211}^{\sigma-\sigma}),
\label{Green_function}
\end{eqnarray}

where operators $G_{2}^{-1}$ and $G_{1}^{-1}$ have the form:

\begin{eqnarray}
G_{2}^{-1}&=&i\frac{\partial}{\partial t}+\xi+i\gamma,\nonumber\\
G_{1}^{-1}&=&i\frac{\partial}{\partial t}-\xi+i\gamma.
\end{eqnarray}

Solution of the Eq. (\ref{Green_function}) describes localized
charge relaxation and consists of the two parts: the first one is
the general solution of the homogeneous equation $n_{1}^{h}(t)$
(right hand part is equal to zero) and the second one is the partial
solution of the inhomogeneous equation $\widetilde{n}_{1}(t)$.

\begin{eqnarray}
              \label{P}
n_1(t)&=&n_{1}^{h}(t)+\widetilde{n}_{1}(t)\nonumber\\&=&n_{1}^{h}(t)+\int_{0}^{t}\textit{\textbf{G}}(t-t^{'})P(t^{'})dt^{'},
\end{eqnarray}

where $\textit{\textbf{G}}(t-t^{'})$- is the Green function of the
Eq. (\ref{Green_function}) with $\delta(t-t^{'})$ in the right hand
part, and $P(t^{'})$-is the right hand part of the Eq.
(\ref{Green_function}), which appears due to the Coulomb
correlations.

General solution of the homogeneous equation has the form:
\cite{Pump}

\begin{eqnarray}
n_{1}^{h}(t)&=&n_{1}^{0}[A^{'}e^{-i(E_{1}-E_{1}^{*})t}\nonumber\\&+&2Re(B^{'}e^{-i(E_{1}-E_{2}^{*})t})
+C^{'}e^{-i(E_{2}-E_{2}^{*})t}], \label{filling_numbers_1}
\end{eqnarray}

where coefficients $A'$, $B'$ and $C'$ are determined as:

\begin{eqnarray}
A^{'}&=&\frac{|E_{2}-\varepsilon_1|^{2}}{|E_{2}-E_{1}|^{2}};\quad
C^{'}=\frac{|E_{1}-\varepsilon_1|^{2}}{|E_{2}-E_{1}|^{2}};\nonumber\\
B^{'}&=&-\frac{(E_{2}-\varepsilon_1)(E_{1}^{*}-\varepsilon_1)}{|E_{2}-E_{1}|^{2}}.
\label{p1}
\end{eqnarray}

Eigenfrequencies $E_{i}$ can be found from the equation:

\begin{eqnarray}
(E-\varepsilon_1)(E-\varepsilon_2+i\gamma)-T^{2}=0,
\end{eqnarray}
and have the form
\begin{eqnarray}
E_{1,2}=\frac{1}{2}(\varepsilon_1+\varepsilon_2-i\gamma)\nonumber\\\pm
\frac{1}{2}\sqrt{(\varepsilon_1-\varepsilon_2+i\gamma)^{2}+4T^{2}}.
\end{eqnarray}

Green function $\textit{\textbf{G}}(t-t^{'})$ of the Eq.
(\ref{Green_function}) can be written as:

\begin{eqnarray}
\textit{\textbf{G}}(t-t^{'})=\sum_{i=1}^{4}a_i
e^{-i\lambda_i(t-t^{'})}\Theta(t-t^{'}),
\end{eqnarray}

where $\lambda_i$-are the roots of the characteristic equation
arising from Eq. (\ref{Green_function}) :

\begin{eqnarray}
\lambda_{1,2}&=&-i\gamma\pm[\frac{4T^{2}+\xi^{2}-\gamma^{2}}{2}\nonumber\\&+&\frac{1}{2}\sqrt{(4T^{2}+\xi^{2}-\gamma^{2})^{2}+4\xi^{2}\gamma^{2}}]^{1/2},\nonumber\\
\lambda_{3,4}&=&-i\gamma\pm
[\frac{4T^{2}+\xi^{2}-\gamma^{2}}{2}\nonumber\\&-&\frac{1}{2}\sqrt{(4T^{2}+\xi^{2}-\gamma^{2})^{2}+4\xi^{2}\gamma^{2}}]^{1/2},
\end{eqnarray}

these roots are connected with the eigenfrequencies $E_{i}$ by the
relations
\begin{eqnarray}
\lambda_{1,2}&=&E_{1,2}-E_{1,2}^{*},\nonumber\\
\lambda_{3}&=&E_{1}-E_{2}^{*},\nonumber\\
\lambda_{4}&=&E_{2}-E_{1}^{*}.
\end{eqnarray}

\begin{figure} [t]
\includegraphics[width=60mm]{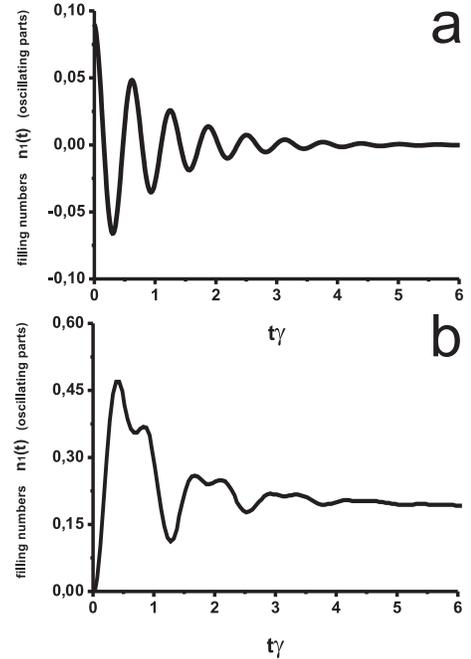}
\caption{Oscillating terms for the two different time evolution
regimes of the filling numbers $n_{1}(t)$. a). $\xi/\gamma=0$; b).
$(\xi+U)/\gamma=0$. Parameters $T/\gamma=0.6$, $\gamma=1$,
$U/\gamma=10$ are the same for all the figures.}
\label{figure_3a_3b}
\end{figure}

Coefficients $a_{i}$ are determined as:

\begin{eqnarray}
a_{1}=\frac{1}{(\lambda_2-\lambda_1)(\lambda_3-\lambda_1)(\lambda_4-\lambda_1)},\nonumber\\
a_{2}=\frac{1}{(\lambda_1-\lambda_2)(\lambda_3-\lambda_2)(\lambda_4-\lambda_2)},\nonumber\\
a_{3}=\frac{1}{(\lambda_1-\lambda_3)(\lambda_2-\lambda_3)(\lambda_4-\lambda_3)},\nonumber\\
a_{4}=\frac{1}{(\lambda_1-\lambda_4)(\lambda_2-\lambda_4)(\lambda_3-\lambda_4)}.
\end{eqnarray}

Let us now focus on the two limit cases when the expressions which
determine the dynamics of the filling numbers have a rather compact
form. The first one corresponds to the situation when the detuning
between the empty energy levels in the QDs is equal to zero:
$\xi/\gamma \ll 1$. The second one deals with the situation when the
sum of the detuning and the half value of Coulomb interaction is
equal to zero. This means that the resonance between the half
occupied energy level in the first QD and the empty level in the
second QD takes place: $(\xi+U)/\gamma\ll 1$. We shall also consider
that in both cases the condition $T\ll\gamma\ll U$ is fulfilled.

\subsection{$\xi/\gamma\ll 1$}

The eigenvalues of the characteristic equation in the first case
($\xi/\gamma\ll 1$) within the accuracy $\frac{T^{2}}{U^{2}}$ have
the form:

\begin{eqnarray}
\lambda_1&=&U-i\frac{2T^{2}\gamma}{U^{2}},\nonumber\\
\lambda_2&=&-i\gamma-i\frac{2T^{2}}{\gamma},\nonumber\\
\lambda_3&=&-2i\gamma-i\frac{2T^{2}}{\gamma},\nonumber\\
\lambda_4&=&-i\gamma.
\end{eqnarray}

So, the evolution operators can be written as:

\begin{eqnarray}
\Phi_{12}(t)&=&\frac{T}{U}(e^{-iUt-\frac{2T^{2}\gamma}{U^{2}}
t}-e^{-\gamma t-\frac{2T^{2}}{\gamma}t}),\nonumber\\
\Phi_{22}(t)&=& (1-\frac{2T^{2}}{U^{2}})
e^{-iUt-\frac{2T^{2}\gamma}{U^{2}} t}+\frac{2T^{2}}{U^{2}}
e^{-\gamma
t-\frac{2T^{2}}{\gamma}t},\nonumber\\
\end{eqnarray}

and time dependence of the pair correlators
$K_{2111}^{\sigma-\sigma}$ and $K_{1211}^{\sigma-\sigma}$ is
determined by the product:

\begin{eqnarray}
K_{2111}^{\sigma-\sigma}(t)&=&\Phi_{12}(t)\Phi_{22}^*(t), \nonumber \\
K_{1211}^{\sigma-\sigma}(t)&=&(K_{2111}^{\sigma-\sigma})^*.
\end{eqnarray}

Expression for the $P(t)$ [see Eq. (\ref{P})] in the case of the
resonance between empty levels $\xi/\gamma=0$ has the form:

\begin{eqnarray}
P(t)=4T^{2}\gamma e^{-\frac{4T^{2}\eta}{\gamma}t}+2T^{2}Ue^{-\gamma
t}\cos(Ut),
\end{eqnarray}

where $\eta=\frac{\gamma^{2}}{U^{2}+\gamma^{2}}$. For $\eta=1/2$ the
inhomogeneous part of the time evolution of the filling numbers
$\widetilde{n}_{1}(t)$ can be written as:

\begin{eqnarray}
\widetilde{n}_{1}(t)&=&\frac{T^{2}}{\gamma^{2}}[(-2\gamma
t-e^{-\frac{2T^{2}}{\gamma}t})
e^{-\frac{2T^{2}}{\gamma}t}+e^{-2\gamma
t}\nonumber\\&+&4e^{-\frac{2T^{2}}{\gamma}t}-4e^{-\gamma t}]+
\frac{2T^{2}Ue^{-\gamma
t}}{\gamma^{3}}[cos(Ut)-1]\nonumber\\&+&0(\frac{T^{2}}{U^{2}}\frac{\gamma}{U}).
\label{1}
\end{eqnarray}

For $\eta\ll1$ the time evolution of the filling numbers
$\widetilde{n}_{1}(t)$ can be determined by:

\begin{eqnarray}
\widetilde{n}_{1}(t)=\frac{1}{1-2\eta}(e^{-4\frac{T^{2}\eta}{\gamma}t}-e^{-\frac{2T^{2}}{\gamma}t})+0(\frac{T^{2}}{\gamma^{2}}).
\end{eqnarray}

\subsection{$(\xi+U)/\gamma\ll 1$}

In the second case of interest ($(\xi+U)/\gamma\ll 1$ but $U/\gamma
\gg 1$) eigenvalues are:

\begin{eqnarray}
\lambda_1&=&\frac{U}{2}-i\frac{8T^{2}\gamma}{U^{2}},\nonumber\\
\lambda_2&=&-i\gamma+\frac{8T^{2}}{U},\nonumber\\
\lambda_3&=&-2i\gamma+\frac{U}{2}-\frac{i8T^{2}\gamma}{U^{2}}+\frac{4T^{2}}{U},\nonumber\\
\lambda_4&=&-i\gamma.
\end{eqnarray}

Evolution operators have the following form:

\begin{eqnarray}
\Phi_{12}(t)&=&\frac{2T}{U}(e^{-i\frac{U}{2}t-\frac{8T^{2}\gamma}{U^{2}}
t}-e^{-\gamma t}),\nonumber\\
\Phi_{22}(t)&=&(1-\frac{8T^{2}}{U^{2}})e^{-i\frac{U}{2}t-\frac{8T^{2}\gamma}{U^{2}}
t}+\frac{8T^{2}}{U^{2}}e^{-\gamma t}.
\end{eqnarray}

When the condition $(\xi+U)/\gamma=0$ is fulfilled, $P(t)$ within
the accuracy $\frac{T^{3}}{U^{3}}$ and $\frac{\gamma^2}{U^2}$ is
determined by the expression:

\begin{eqnarray}
P(t)=-T^{2}U^2(e^{i\frac{U}{2}t-\gamma
t}+h.c.)-4T^{2}\gamma^{2} e^{-\frac{16T^{2}\gamma}{U^{2}}t}.\nonumber\\
\end{eqnarray}

The inhomogeneous part of the time evolution of the filling numbers
$\widetilde{n}_{1}(t)$ with the accuracy $\frac{T^{2}}{U^{2}}$ has
the form:

\begin{eqnarray}
\label{inhomog}
\widetilde{n}_{1}(t)&=&-\frac{4}{7}(1-e^{-14\frac{T^{2}\gamma}{U^{2}}t})
e^{-\frac{2T^{2}\gamma}{U^{2}}t}\nonumber\\
&-& 2\frac{T^{2}}{\gamma U} e^{-\gamma t}
\sin(\frac{U}{2}t)+0(\frac{T^{2}}{U^{2}}).
\end{eqnarray}

\begin{figure} [t]
\includegraphics[width=60mm]{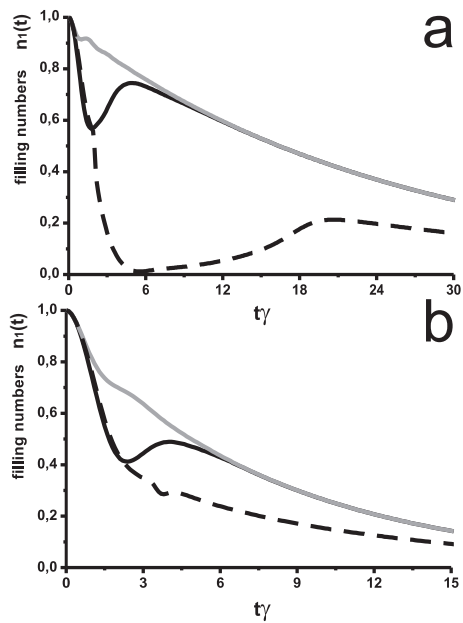}
\caption{Time dependence of the filling numbers for the electrons
$n_{1}(t)$ in the presence of Coulomb interaction: comparison of the
exact solution and the mean-field approximation. Black line
corresponds to the exact solution, black dashed line corresponds to
the mean-field approximation. Grey line demonstrates relaxation of
the localized charge in the absence of Coulomb interaction. a).
$U/\gamma=5$, $\xi/\gamma=-3$; b). $U/\gamma=3$, $\xi/\gamma=-2$.
Parameters $T/\gamma=0.6$, $\gamma=1$ are the same for all the
figures.} \label{figure_4a_4b}
\end{figure}

It is necessary to point out that relaxation of the filling numbers
in the proposed model can be analyzed by means of  more simple
method
--- the self-consistent mean-field approximation.\cite{Anderson, Mantsevich_1} In this approximation correlation functions $U\langle
\hat n_{i}^{-\sigma}\hat n_{ij}^{\sigma}\rangle$ in the Eqs.
(\ref{system}) are substituted by the expressions $U\langle \hat
n_{i}^{-\sigma}\rangle\langle \hat n_{ij}^{\sigma}\rangle$. Such
substitution is valid in the case when filling numbers for the
localized electrons $n_{i}^{-\sigma}$ change their values rather
slow. Calculation scheme consists of the two steps. On the first
step one has to substitute the initial energy level position
$\varepsilon_i$ by the expression
$\widetilde{\varepsilon}_i=\varepsilon_i+U\langle \hat
n_{i}^{-\sigma}\rangle$ and to evaluate the time dependent filling
numbers. The second step deals with the self-consistent calculation
of the time dependent filling numbers for the electrons. For some
ranges of the system parameters mean-field approximation reveals
qualitatively good results. \cite{Mantsevich_1} But in general case
the mean-field approximation is insufficient to describe the
relaxation processes in the system with correlations.

\section{Results and discussion}

Time evolution of the filling numbers for the electrons strongly
depends on the relations between the system parameters.

\begin{figure} [t]
\includegraphics[width=60mm]{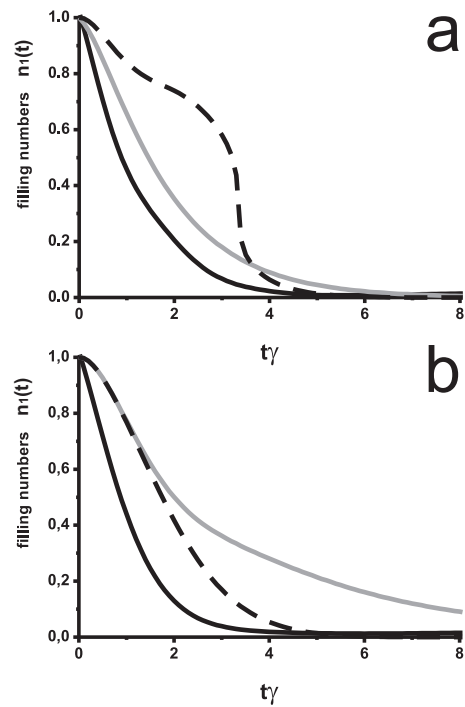}
\caption{Relaxation of the filling numbers $n_{1}(t)$ in the
presence of Coulomb interaction in the case of the resonant
tunneling between the empty energy levels in the QDs. Black line
corresponds to the exact solution, black dashed line corresponds to
the mean-field approximation. Grey line demonstrates relaxation of
the localized charge in the absence of Coulomb interaction. a).
$\xi/\gamma=0$, $U/\gamma=3$; b). $\xi/\gamma=0$, $U/\gamma=1$.
Parameters $T/\gamma=0.6$, $\gamma=1$ are the same for all the
figures.} \label{figure_5a_5b}
\end{figure}

If condition $(\xi+U)/\gamma\ll 1$ is fulfilled, the Coulomb
interaction value increasing leads to the decreasing of the filling
numbers relaxation rate [see Fig. \ref{figure_2a_2c}(a)]. For the
large $U$ relaxation rate is rather slow and is of the order of
$\gamma_{nonres}=2\frac{T^{2}\gamma}{U^{2}}$ which is typical for
the system of two coupled QDs without Coulomb interaction with
$|\xi| \simeq U$. By the decreasing of the Coulomb interaction value
$U$ we achieve the situation of resonant tunneling between the
localized states and consequently relaxation rate becomes larger. On
the Fig. \ref{figure_2a_2c}(c) the situation of resonant tunneling
between the empty energy levels $\xi/\gamma=0$ is demonstrated. In
this case the relaxation of the localized charge takes place with
the typical rate very close to the value
$\gamma_{res}=2\frac{T^{2}}{\gamma}$ and is almost independent on
the Coulomb interaction value. Let us notice that relaxation
processes are governed not only by the typical exponents $e^{-\gamma
t}$ and $e^{-\frac{2T^{2}}{\gamma}t}$ but also by the
pre-exponential factor, which linearly increases in time in the
resonant case [see Eq. (\ref{1})].

A very special relaxation regime exists in the system if condition
$\frac{\xi+U}{\gamma}\sim1$ takes place [see Fig.
\ref{figure_2a_2c}(b)]. In this regime Coulomb correlations result
in formation of a dip in the time evolution of the localized charge.
At the initial relaxation stage the charge in the first QD rapidly
decreases due to the almost resonant relation between the level in
the second QD and effective single electron energy in the first dot.
It follows from the third and the fourth Eqs. (\ref{system}) of the
system that changing of the effective energy levels detuning is
determined by $URe[\frac{\langle
\hat{n}_{1}^{-\sigma}(t)\hat{n}_{12}^{\sigma}(t)\rangle}{\langle
\hat{n}_{12}^{\sigma}(t)\rangle}]$ which differs from the typical
mean-field expression $U\langle \hat{n}_{i}^{-\sigma}(t)\rangle$.
\cite{Anderson}

At a certain instant of time the effective single electron level
falls down beneath the level in the second QD. At this moment the
inverse charge begins to flow from the second QD to the first one.
The occupation in the first QD demonstrates significant increasing
after reaching minima value (the dip formation). Filling numbers
almost reach the initial value for the large values of Coulomb
interaction. After the dip formation the typical time scale which
determines relaxation of the filling numbers is close enough to the
value $\gamma_{nonres}=2\frac{T^{2}\gamma}{\xi^{2}}$. This
explanation gives qualitative picture of the dips formation. The
exact solution shows, that Coulomb correlations are responsible for
such non-monotonic behavior. This effect is determined by the
inhomogeneous part of the exact solution for time evolution of the
filling numbers in the first QD [see the first term in Eq.
(\ref{inhomog})]. And this inhomogeneous part appears due to
complete account for time dependence of the high order correlators
[$P(t)$ in Eq (\ref{Green_function}) and Eq. (\ref{P}))]. That is
why time evolution of the filling numbers for the electrons differs
considerably from mean-field approximation. The width of the dip can
be roughly estimated as $1/8\cdot\gamma_{nonres}^{-1}$.

We would like to stress that the non-monotonic behavior, which we
discussed above,
 is not connected with the usual quantum oscillations between two energy levels.
Such oscillations also take place during time evolution, but the
amplitude of these oscillations is rather small (of the order
$\frac{T^{2}}{U^{2}}$). Only these small oscillating contributions
to the total electron density are shown on the Fig.
\ref{figure_3a_3b}. Oscillations are always present in the case of
strong Coulomb interaction for all the values of the ratio
$T/\gamma$. We found out that besides the oscillations governed by
the system parameters $T$ and $U$, oscillations with the double
period exist in the system. Oscillation period doubling is mostly
pronounced in the case when resonant tunneling takes place between
the half occupied energy level in the first QD with the initial
charge and empty level in the dot coupled with the continuous
spectrum states $(\xi+U)/\gamma\ll 1$ [see Fig.
\ref{figure_3a_3b}(b)]. Double period oscillations disappears with
the decreasing of energy levels detuning $\xi$. In this case
oscillations period is determined by the value of the Coulomb
interaction [see Eq. (\ref{1})].

Comparison between the exact solution and the mean-field
approximation is demonstrated on the Figs.
\ref{figure_4a_4b}-\ref{figure_5a_5b}. It is clearly evident that
both methods reveal such similar peculiarities of the system
behavior as several time ranges with considerably different
relaxation rates. For some ranges of the system parameters formation
of the dip can be also reproduced in the mean-field approximation
(see Fig. \ref{figure_4a_4b}). Figure \ref{figure_4a_4b} also
demonstrates similar behavior of the exact and the mean-field
solutions at the initial stage of relaxation. But the dip reproduces
incorrectly in the mean-field approximation.

\begin{figure*} [t]
\includegraphics[width=170mm]{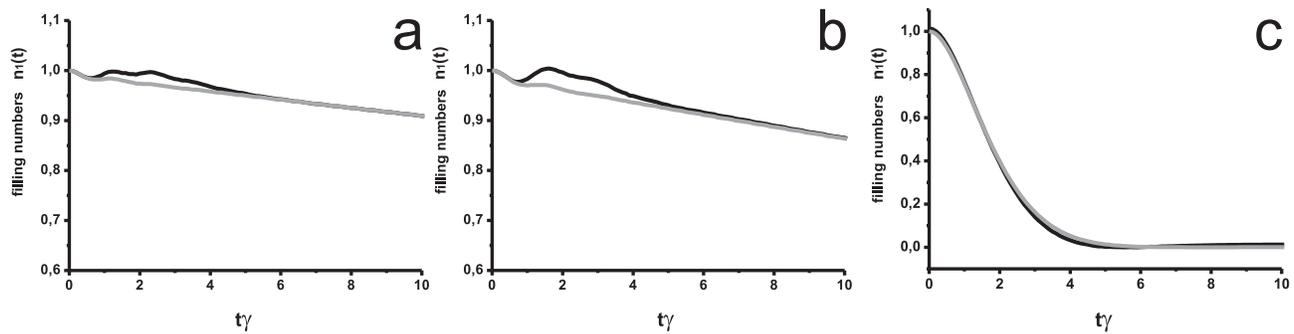}
\caption{The influence of the Coulomb interaction in the second QD
on the time evolution of the filling numbers $n_{1}(t)$ in the first
QD. Black line corresponds to the exact solution. Grey line
demonstrates relaxation of the localized charge in the absence of
Coulomb interaction. a).$\xi/\gamma=-5$; b).$\xi/\gamma=-4$;
c).$\xi/\gamma=0$. Parameters $T/\gamma=0.6$, $\gamma=1$,
$U/\gamma=5$ are the same for all the figures.} \label{figure_6a_6c}
\end{figure*}

In the case of resonant tunneling between the energy levels in the
QDs ($\xi/\gamma = 0$) the exact solution and the mean-field
approximation reveal strong mismatch [see Fig.
\ref{figure_5a_5b}(a)]. Exact solution demonstrates rather smooth
time evolution of the localized charge while the solution obtained
by means of the mean-field approximation reveals abrupt changing of
the localized charge amplitude. If the Coulomb repulsion decreases
the correspondence between the exact and the mean-field solutions
becomes better [see Fig. \ref{figure_5a_5b}(b)]

Finally let us return to the influence of the Coulomb repulsion in
the second QD on the evolution of the filling numbers. In this
situation time evolution of the filling numbers for the electrons
can be analyzed by means of the equations obtained for the model
when Coulomb interaction acts in the first QD [see Eq.
(\ref{system})] after substituting the value $U\hat n_{1}^{-\sigma}$
by the $U\hat n_{2}^{-\sigma}$ in the Eq. (\ref{system}). The
results are shown on the Fig. \ref{figure_6a_6c} and it is clearly
evident that in this case the influence of Coulomb correlations on
the relaxation of the filling numbers is rather weak.

\section{Conclusions}

We have studied time evolution of the filling numbers  in the system
of two interacting QDs coupled with the continuous spectrum states
in the presence of Coulomb interaction in one of the dots for a wide
range of the system parameters. The solution describing the system
dynamics was analyzed in the assumption that the band and localized
filling numbers for the electrons are uncoupled. This solution
exactly takes into account all order correlators for the localized
electrons in the QDs.

We found strongly different relaxation regimes in the system of
coupled QDs depending on the ratios between the system parameters.
Interesting manifestation of  Coulomb correlations is the formation
of the dip in the time evolution of the localized charge. Such
reentrant charge behavior is not the result of simple quantum
oscillations between the two energy levels. Oscillations of this
type are also present in the system but have small amplitude in the
case of the strong Coulomb interaction. Interaction effects lead to
the appearance of oscillations with double period at particular
range of parameters together with the oscillations governed by the
detuning between the energy levels.

We compared our results with the mean-field approximation. The
mean-field approximation can give in some cases qualitatively
similar peculiarities of the system behavior: several time ranges
with considerably different relaxation rates and dip's formation.
But in many regimes the results of the mean-field approximation do
not coincide with the exact solution. Even if the mean-field
approximation qualitatively correctly predicts appearance of the
dip, it's shape and width strongly differs from the exact solution.

\section{ACKNOWLEDGMENTS}

This work was partly supported by the RFBR, Leading Scientific
School grants and Russian Ministry of Science and Education
programs.


\pagebreak

\end{document}